\documentclass{article}
\usepackage{algorithm,algorithmic}
\usepackage{graphicx}
\usepackage{amsmath}
\usepackage{amssymb}
\usepackage{amsthm}
\usepackage{bbm}

\newtheorem{definition}{Definition}
\newtheorem{theorem}{Theorem}
\newtheorem{example}{Example}

\begin{document}
\title{Reasoning about Games via a First-order Modal Model Checking Approach}
\author{Davi Romero de Vasconcelos \footnote{Universidade Federal do
Cear\'a (UFC), Quixad\'a, CE, Brazil }
\and
Edward Hermann Haeusler\footnote{Departamento
de Inform\'atica, Pontif\'icia Universidade Cat\'olica do Rio de
Janeiro (PUC-Rio), Rio de Janeiro-RJ, Brazil}}
\date{}
\maketitle

\begin{abstract}
In this work, we
present a logic based on first-order CTL, namely Game Analysis Logic
(GAL), in order to reason about games. We relate models and solution
concepts of Game Theory as models and formulas of GAL, respectively.
Precisely, we express extensive games with perfect information as
models of GAL, and Nash equilibrium and subgame perfect equilibrium
by means of formulas of GAL. From a practical point of view, we
provide a GAL model checker in order to analyze games automatically.
We use our model checker in at least two directions: to find
solution concepts of Game Theory; and, to analyze players that are
based on standard algorithms of the AI community, such as the
minimax procedure.
\end{abstract}

\section{Introduction}
Games are abstract models of decision-making in which
decision-makers (players) interact in a shared environment to
accomplish their goals. Several models have been proposed to analyze
a wide variety of applications in many disciplines such as
mathematics, computer science and even political and social sciences
among others.

Game Theory \cite{Games} has its roots in the work of von Neumann
and Morgenstern \cite{TGEB} and uses mathematics in order to model
and analyze games in which the decision-makers pursue rational
behavior in the sense that they choose their actions after some
process of optimization and take into account their knowledge or
expectations of the other players' behavior. Game Theory provides
general game definitions as well as reasonable solution concepts for
many kinds of situations in games. Typical examples of this kind of
research come from phenomena emerging from Markets, Auctions and
Elections.

Although historically Game Theory has been considered more suitable
to perform quantitative analysis than qualitative ones, there has
been a lot of approaches that emphasizes Game Analysis on a
qualitative basis, by using an adequate logic in order to express
games as well as their solution concepts. Some of the most
representatives of these logics are: Coalitional Logic
\cite{PaulyThesis01}; Alternating-time Temporal Logic (ATL)
\cite{ATL} and its variation Counter-factual ATL (CATL)
\cite{WiebeCATL}; Game Logic \cite{GameLogicPaulyP03a}; Game Logic
with Preferences \cite{PreferencesGameLogicsOtterlooHW04};
Coalitional Game Logic (CGL) \cite{WiebeCGL} that reasons about
coalitional games. To see more details about the connections and
open problems between logic and games, we point out
\cite{OpenProblemsLogicGameVanBenthem}.

The technique of Model Checking \cite{SMV} is frequently employed in
computer science to accomplish formal validation of both software
and hardware \cite{Burck91}. Model Checking consists of achieving
automatic verification and other forms of formal analysis of a
system behavior. A lot of implementations are available in the
literature, such as Symbolic Model Verifier (SMV) \cite{Mc93Thesis},
SPIN \cite{Hol97} and MOCHA \cite{Mocha}. Some other implementations
also include specific features in the modeling, UPPAAL \cite{Uppaal}
works in real-time, HYTECH \cite{Hytech} with hybrid automata and
PRISM \cite{Parker02} with stochastic automata. Recently, model
checking has also been used to verify proprieties in games
\cite{Vasconcelos03,CheckersSMC01,PreferencesGameLogicsOtterlooHW04,VerificationGamesAAMAS06}.

There is a wide range of problems approachable by means of Game
Theory. Besides problems and models coming from economics, which
usually have quantitative features, as those normally present in
econometric models, there is also a range of problems that are
strongly related to Multi-Agent systems modeling and that can be
consequently also  validated by means of well-known CAV tools.
However, the presence of intrinsic and quantitative measures in this
kind of modeling  prevent us from an standard use of the most
popular (and efficient) CAV tools, such as Model Checkers (MCs) based on
the propositional logic language.  One could argue that MCs, like
the SPIN MC, that have a richer operational semantics, such as its
ability to assign computable meaning to transitions by means of
fragments of a Programming Language like coding assertion, might be
the right answer to the specification of such kind of modeling.
However, the SPIN logic language is not a First-Order logic
language, and hence,  cannot make assertions on the internal
structure of an state, mainly regarding the relationship between the
values assigned to the individuals in these states (worlds),
properties regarding the very individuals and so generalizations and
existential assertions on a state and its individuals cannot be, in
general, expressible.  Most of the solution concepts used in Game
Theory are expressed as general assertions on the relationship
between individuals of a possible state-of-affairs in the game.  The
SPIN logic language is unable, in general,  to express such kind of
concept either. Concerning the usefulness of an approach based on a
logic language more expressible than the presently used in CAV
tools, it is worth mentioning new directions in the MC community
towards the use of First-Order Logic (see \cite{CADE-17,MC-FOL-01}).
Thus, this article contributes for this kind of research in the
Formal Methods community, by providing a First-Order based approach
to the problem of validating models able to be expressed by Game
Theoretical means. The present approach is not so restricted, since
the authors have already presented  a result showing how Multi-Agent
systems can be viewed and strongly considered as a Game, such that,
main solution concepts on the Game side represent important concepts
on the Systems side (\cite{VasconcelosWRAC}).

The aim of this article is to present GAL (Game Analysis Logic), a
logic based on first-order CTL, in order to reason about games in
which a model of GAL is a game, and a formula of GAL is an analysis.
We illustrate our approach by showing that GAL is suitable to
express models of Game Theory as well as their solution concepts.
Precisely, we specify extensive game with perfect information by
means of models of GAL. We also express their main solution concepts
- namely Nash equilibrium and subgame perfect equilibrium - by means
of formulas of GAL. In \cite{VasconcelosThesis07}, we express the
standard noncooperative models (strategic games and the solution
concept of Nash equilibrium) and cooperative models (coalition game
and the solution concept of Core). In this article, we focus on the
extensive games and the solution concepts of Nash equilibrium and
subgame perfect equilibrium.

As  GAL has a first-order apparatus, we are able to define many
concepts, such as utility, in an easier way, when compared to the
logics mentioned above. Moreover, a first-order apparatus is
essential to model and reason about social problems that have been
modeled by Game Theory, Econometric Models, etc, as already said. It
is worth mentioning that the ATL logic, in which the operators of
CTL are parameterized by sets of players, can be seen as a fragment
of GAL, using the first-order feature of GAL; thus, there is no need
for such a parameterization in GAL. In addition, the CGL logic,
which is designed to reason about cooperative models, can also be
embed in GAL. See \cite{VasconcelosThesis07} for the proofs that ATL
and CGL can be seen as fragments of GAL. We do not focus on such
proofs here.

We also provide a model checking algorithm for GAL in order to
demonstrate that GAL can be used in practice to analyze games
automatically. We have a prototype of a model checker for GAL that
has been developed according to the main intentions of the approach
advocated here. The model checker is available for download at
www.tecmf.inf.puc-rio.br/DaviRomero. All of the examples in this
article are implemented in the tool. We will show that, using our
prototype, we are able to find solution concepts of Game Theory and
to analyze players that are based on standard algorithms
\cite{Russell02} of the AI community.

This work is divided into six parts: Section 2 introduces Game
Analysis Logic; A model checking algorithm for GAL is presented in
Section 3. Standard concepts of Game Theory are expressed in GAL in
Section 4. Section 5 presents some experimental results using our
algorithm. Finally, Section 6 concludes this work.

\section{Game Analysis Logic (GAL)}
GAL is a many-sorted modal first-order logic language that is a
logic based on the standard Computation Tree Logic (CTL)
\cite{Clarke81}. A game is a model of GAL, called game analysis
logic structure, and an analysis is a formula of GAL.

The \emph{games} that we model are represented by a set of states
$\mathcal{S}E$ and a set of actions $\mathcal{CA}$.

A \emph{state} is defined by both a first-order interpretation and a
set of players, where: 1- The first-order interpretation is used to
represent the choices and the consequences of the players'
decisions. For example, we can use a list to represent the history
of the players' choices until certain state; 2- The set of players
represents the players that have to decide simultaneously at a
state. This set must be a subset of the players' set of the game.
The other players cannot make a choice at this state. For instance,
we can model games such as auction games, where all players are in
all states, or even games as Chess or turn-based synchronous game
structure, where only a single player has to make a choice at each
state. Notice that we may even have some states where none of the
players can make a decision that can be seen as states of the
nature.

An \emph{action} is a relation between two states $e_{1}$ and
$e_{2}$, where all players in the state $e_{1}$ have committed
themselves to move to the state $e_{2}$. Note that this is an
extensional view of how the players committed themselves to take a
joint action.

We refer to $(A_{k})_{k\in K}$ as a sequence of $A_{k}$'s with the
index $k\in K$. Sometimes we will use more than one index as in the
example $(A_{k,l})_{k,l\in K\times L}$. We can also use
$(A_{k},B_{l})_{k\in K, l\in L}$ to denote the sequence of
$(A_{k})_{k\in K}$ followed by the sequence $(B_{l})_{l\in L}$.
Throughout of this article, when the sets of indexes are clear in
the context, we will omit them.

A \emph{path} $\pi(e)$ is a sequence of states (finite or infinite)
that could be reached through the set of actions from a given state
$e$ that has the following properties: 1- The first element of the
sequence is $e$; 2- If the sequence is infinite
$\pi(e)=(e_{k})_{k\in\mathbb{N}}$, then $\forall k\geq0$ we have
$\langle e_{k} ,e_{k+1}\rangle\in\mathcal{CA}$; 3- If the sequence
is finite $\pi(e)=(e_{0},\ldots,e_{l})$, then $\forall k$ such that
$0\leq k<l$ we have $\langle e_{k},e_{k+1}\rangle\in\mathcal{CA}$
and there is no $e^{\prime}$ such that $\langle
e_{l},e^{\prime}\rangle\in\mathcal{CA}$. The game behavior is
characterized by its paths that can be finite or infinite. Finite
paths end in a state where the game is over, while infinite ones
represent a game that will never end.

Below we present the formal syntax and semantics of GAL. As usual,
we call the sets of sorts $S$, predicate symbols $P$, function
symbols $F$ and players $N$ as a non-logic language in contrast to
the logic language that contains the quantifiers and the
connectives. We define a term of a sort in a standard way. We denote
a term $t$ of sort $s$ as $t_{s}$. The modalities can be read as
follows.
\begin{itemize}
\item $[EX]\alpha$ - `exists a path $\alpha$ in the next state'
\item $[AX]\alpha$ - `for all paths $\alpha$ in the next
state'
\item $[EF]\alpha$ - `exists a path $\alpha$ in the future'
\item $[AF]\alpha$ - `for all paths $\alpha$ in the future'
\item $[EG]\alpha$ - `exists a path $\alpha$ globally'
\item $[AG]\alpha$ - `for all paths $\alpha$ globally'
\item $E(\alpha\mathcal{U}$$\beta)$ - `exists a path
$\alpha$ until $\beta$'
\item $A(\alpha\mathcal{U}$$\beta)$ - `for all paths $\alpha$ until $\beta$'
\end{itemize}

\begin{definition}[Syntax of GAL]
Let $\langle S,F,P,N\rangle $ be a non-logic language, and
$t_{s_{1}}^{1},...,t_{s_{n}}^{n}$ be terms, and $t_{s_{1}}^{\prime}$
be a term, and $p:s_{1}...s_{n}$ be a predicate symbol, and $i$ be a
player, and $x_{s}$ be a variable of sort $s$. The \textbf{logic
language of GAL} is generated by the following BNF definition:
 \[\Phi::=
\top~|~\bot~|~i~|~p(t_{s_{1}}^{1},\ldots,t_{s_{n}}^{n})~|~(t_{s_{1}}^{1}\approx
t_{s_{1}}^{\prime})~|~(\lnot\Phi)~|~(\Phi\wedge\Phi)~|~(\Phi\vee\Phi)~|~(\Phi\rightarrow\Phi)\]
\[|~[EX]\Phi~|~[AX]\Phi~|~[EF]\Phi~|~[AF]\Phi~|~[EG]\Phi~|~[AG]\Phi~|~E(\Phi~\mathcal{U}~\Phi)~|~A(\Phi~\mathcal{U}~\Phi)\]
\[~|~\exists x_s\Phi~|~\forall x_s\Phi\]
\end{definition}
It is well-known that the operators
$\wedge,\vee,\bot,[EX],[AF],[EF],[AG],[EG]$ and $\forall x$ can be
given by the following usual abbreviations.
\begin{itemize}
\item
$\bot$ $\Longleftrightarrow$ $\lnot\top$
\item $\alpha\wedge\beta$  $\Longleftrightarrow$
$\lnot(\alpha\rightarrow\lnot\beta) $
\item $\alpha\vee\beta$
$\Longleftrightarrow$ $(\lnot\alpha\rightarrow\beta)$
\item $[EX]\alpha$ $\Longleftrightarrow$ $\lnot[AX]\lnot\alpha$
\item $[AF]\alpha$ $\Longleftrightarrow$
$A(\top~\mathcal{U}~\alpha)$
\item $[EF]\alpha$
$\Longleftrightarrow$ $E(\top~\mathcal{U}~\alpha)$
\item $[AG]\alpha$ $\Longleftrightarrow$ $\lnot
E(\top~\mathcal{U}~\lnot\alpha)$
\item $[EG]\alpha$
$\Longleftrightarrow$ $\lnot A(\top~\mathcal{U}~\lnot\alpha)$
\item $\forall x\alpha(x)$ $\Longleftrightarrow$ $\lnot\exists
x\lnot\alpha(x)$
\end{itemize}
\begin{definition}[Structure of GAL]
Let $\langle S,F,P,N\rangle $ be a non-logic language of GAL. A
\textbf{Game Analysis Logic Structure} for this non-logic language
is a tuple $\mathcal{G}=\langle
\mathcal{S}E,\mathcal{S}E_{o},\mathcal{CA},~(\mathcal{D}_{s}),$
$~(\mathcal{F}_{f,e}),~(\mathcal{P}_{p,e}),~(N_{e})\rangle$ such
that:
\begin{itemize}
\item $\mathcal{S}E$ is a non-empty set, called the set of
states.
\item $\mathcal{S}E_{o}$ is a set of initial states, where
$\mathcal{S}E_{o}\subseteq\mathcal{S}E$.
\item For each state $e\in\mathcal{S}E$, $N_{e}$ is a subset of
$N$.
\item $\mathcal{CA}\subseteq\mathcal{S}E\times\mathcal{S}E$,
called the set of actions of the game\footnote{This relation is not
required to be total as in the CTL case. The idea is because we have
finite games.}, in which if there is at least one player in the
state $e_{1}$, then exists a state $e_{2}$ such that $\langle
e_{1},e_{2}\rangle\in\mathcal{CA}$.
\item For each sort $s\in S$, $\mathcal{D}_{s}$ is a non-empty
set, called the domain of sort $s$\footnote{In algebraic terminology
$\mathcal{D}_{s}$ is a carrier for the sort $s$.}.
\item For each function symbol $f:s_1\times\ldots\times s_n\rightarrow s$ of $F$ and each
state $e\in \mathcal{S}E$, $\mathcal{F}_{f,e}$ is a function such
that $\mathcal{F}_{f,e}:\mathcal{D}_{s_{1}}\times\ldots\times
\mathcal{D}_{s_{n}}\rightarrow \mathcal{D}_{s}$.
\item For each predicate symbol $p:s_1\times\ldots\times s_n$ of $P$ and state $e\in
\mathcal{S}E$, $\mathcal{P}_{p,e}$ is a relation such that
$\mathcal{P}_{p,e}\subseteq \mathcal{D}_{s_{1}}\times \ldots\times
\mathcal{D}_{s_{n}}$.
\end{itemize}
\end{definition}

A \textbf{function or predicate is rigidly interpreted} if its
interpretation is the same for every state. A \textbf{GAL-structure
is finite} if the set of states $\mathcal{S}E$ and each set of
domains $D_{s}$ are finite. Otherwise, it is infinite. Note that
even when a GAL-structure is finite we might have infinite paths.

In order to provide the semantics of GAL, we define a valuation
function as a mapping $\sigma_{s}$ that assigns to each free
variable $x_{s}$ of sort $s$ some member $\sigma_{s}(x_{s})$ of
domain $\mathcal{D}_{s}$. As we use terms, we extend every function
$\sigma_{s}$ to a function $\bar{\sigma}_{s}$ from state and term to
element of sort $s$ that is done in a standard way. When the
valuation functions are not necessary, we will omit them.

\begin{figure}[h]
 \begin{tabular}[l]{ccc}
  \raisebox{-0pt}{
      \includegraphics[width=.27\textwidth]{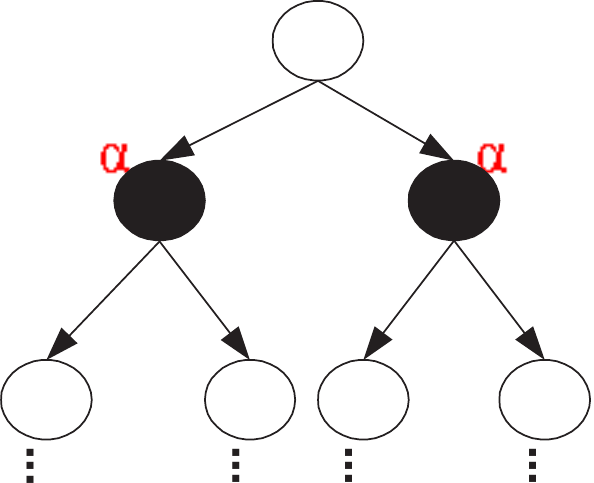}
      }
  &
  \raisebox{-0pt}{
    \includegraphics[width=.27\textwidth]{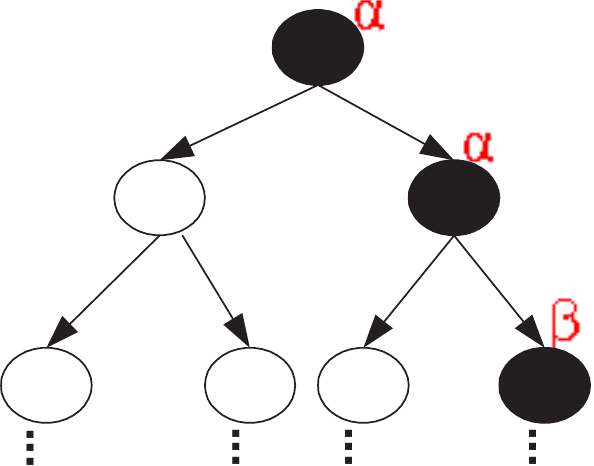}
    }
  &
  \raisebox{-0pt}{
    \includegraphics[width=.27\textwidth]{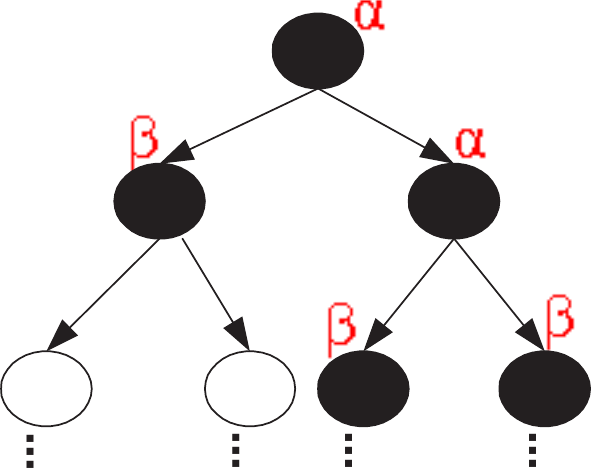} 
    } \\
    (a) - $[AX]\alpha$ & (b) - $E(\alpha~\mathcal{U}\beta)$ & (c) - $A(\alpha~\mathcal{U}\beta)$
\end{tabular}
\caption{Modal Connectives of GAL.}\label{figModalConectives}
\end{figure}

\begin{definition}[Semantics of GAL]
Let
$\mathcal{G}=\langle\mathcal{S}E,\mathcal{S}E_{o},\mathcal{CA},(\mathcal{D}_{s}),
(\mathcal{F}_{f,e}),(\mathcal{P}_{p,e}),$ $(N_{e})\rangle $ be a
GAL-structure, and $(\sigma_{s})$ be valuation functions, and
$\alpha$ be a GAL-formula, where $s\in S, f\in F, p\in P$ and
$e\in\mathcal{S}E$. \textbf{We write
$\mathcal{G},(\sigma_{s})\models_{e}\alpha$ to indicate that the
state $e$ satisfies the formula $\alpha$ in the structure
$\mathcal{G}$ with valuation functions $(\sigma_{s})$}. The formal
definition of satisfaction $\models$ proceeds as follows:
\begin{itemize}
\item $\mathcal{G},(\sigma_{s})\models_{e}\top$.
\item $\mathcal{G},(\sigma_{s})\models_{e}i\Longleftrightarrow
i\in N_{e}$
\item $\mathcal{G},(\sigma_{s})\models_{e}p(t^{1}_{s_{1}},...,t^{n}_{s_{n}}
)\Longleftrightarrow\langle
\bar{\sigma}_{s_{1}}(e,t^{1}_{s_{1}}),...,\bar{\sigma}_{s_{n}}
(e,t^{n}_{s_{n}})\rangle \in \mathcal{P}_{p,e}$
\item $\mathcal{G},(\sigma_{s})\models_{e}(t^{1}_{s_{1}}\approx
t^{\prime}_{s_{1}})\Longleftrightarrow
\bar{\sigma}_{s_1}(e,t^{1}_{s_1})=\bar{\sigma}_{s_1}(e,t^{\prime}_{s_1})$
\item  $\mathcal{G},(\sigma_{s})\models_{e}\lnot\alpha$
$\Longleftrightarrow$ NOT
$\mathcal{G},(\sigma_{s})\models_{e}\alpha$
\item  $\mathcal{G},(\sigma_{s})\models_{e}(\alpha\rightarrow\beta)$
$\Longleftrightarrow$ IF $\mathcal{G},(\sigma_{s})\models_{e}\alpha$
THEN $\mathcal{G},(\sigma_{s})\models_{e}\beta$
\item  $\mathcal{G},(\sigma_{s})\models_{e}[AX]\alpha\Longleftrightarrow$
$\forall e^{\prime}\in\mathcal{S}E\ $such that $\langle
e,e^{\prime}\rangle \in\mathcal{CA}$ we have
$\mathcal{G},(\sigma_{s})\models_{e^{\prime}}\alpha$ (see Figure
\ref{figModalConectives}.a).
\item $\mathcal{G},(\sigma_{s})\models_{e}E(\alpha~\mathcal{U}$
$\beta)$ $\Longleftrightarrow$ exists a finite (or infinite) path
$\pi(e)=(e_{0}e_{1}e_{2}...e_{i}),$ such that exists a $k$ where
$k\geq0$, and
 $\mathcal{G},(\sigma_{s})\models_{e_{k}}\beta$, and for all $j$ where $0\leq j<
k$, and $\mathcal{G},(\sigma_{s})\models_{e_{j}}\alpha$ (see Figure
\ref{figModalConectives}.b).
\item $\mathcal{G},(\sigma_{s})\models_{e} A(\alpha~\mathcal{U}$ $\beta)$
$\Longleftrightarrow$ for all finite (and infinite) paths such that
$\pi(e)=(e_{0}e_{1}e_{2}...e_{i}),$ exists a $k$ where $k\geq0$, and
 $\mathcal{G},(\sigma_{s})\models_{e_{k}}\beta$, and for all $j$ where $0\leq j<
k$, and $\mathcal{G},(\sigma_{s})\models_{e_{j}}\alpha$ (see Figure
\ref{figModalConectives}.c).
\item  $\mathcal{G},(\sigma_{s},\sigma_{s_{k}})\models_{e}\exists
x_{s_{k}}\alpha\Longleftrightarrow$ exists $d\in
\mathcal{D}_{s_{k}}$ such that
$\mathcal{G},(\sigma_{s},\sigma_{s_{k}}(x_{s_{k}}|d))\models_{e}\alpha$,
where $\sigma_{s_{k}}(x_{s_{k}}|d)$ is the function which is exactly
like $\sigma_{s_{k}}$ except for one thing: At the variable
$x_{s_{k}}$ it assumes the value $d$. This can be expressed by the
equation:
\[ \sigma_{s}(x_{s_{k}}|d)(y)=\left\{
\begin{array}
[c]{l} \sigma_{s}(y),  \textrm{ if }\ y\neq x_{s_{k}} \\ d, \qquad
\textrm{ if }\ y=x_{s_{k}}
\end{array}
\right. \]
\end{itemize}
\end{definition}

\section{Satisfatibility and Model Checking for GAL}
\label{sectionGALV} It is well-known that there is no sound and
complete system for a first-order CTL
\cite{UndecidableFOCTLMontagnaPT02}. Thus, GAL is also
non-axiomatizable. However, we argue that, using model checking for
GAL, we can reason about games as well. Besides that we can also
define a non-complete axiomatization of GAL in order to cope with
proofs of interesting results, such as the existence of mixed Nash
equilibrium in strategic games, but we do not focus on this in this
article. In the sequel we state the model checking problem for GAL
and also discuss briefly a model checking algorithm for GAL.

Let $\mathcal{G}=\langle
\mathcal{S}E,\mathcal{S}E_{o},\mathcal{CA},(\mathcal{D}_{s}),$
$(\mathcal{F}_{f,e}),(\mathcal{P}_{p,e}),(N_{e})\rangle $ be a
GAL-structure with the non-logic language $\langle S,F,P,N\rangle $,
and $(\sigma_{s})$ be valuation functions and $\alpha$ be a
GAL-formula. The GAL model checking problem is to find the set of
states that satisfies the formula $\alpha$.
\[\{e\in\mathcal{S}E \textrm{ }| \textrm{ }\mathcal{G},(\sigma_{s})\models_{e}\alpha\}\]

In order to have a model checking algorithm for GAL, we assume that
all of the games are finite; however, we might still have infinite
behavior.

The algorithm for solving the GAL model checking problem uses an
explicit representation of the GAL-structure as a labelled, directed
graph. The nodes represent the states $\mathcal{S}E$, the arcs in
the graph provide the set of actions $\mathcal{CA}$ and the labels
associated with the nodes describe both the players' set $N_{e}$ and
the first-order interpretation (the interpreted functions' set
$(\mathcal{F}_{f,e})$ and the interpreted predicates' set
$(\mathcal{P}_{p,e})$). The algorithm also uses the functions
$\mathcal{D}:S\rightarrow \mathcal{D}_{s}$,
$\mathcal{N}:\mathcal{S}E\rightarrow N_{e}$,
$\mathcal{F}:F\times\mathcal{S}E\rightarrow\mathcal{F}_{f,e}$ and
$\mathcal{P}:P\times\mathcal{S}E\rightarrow\mathcal{P}_{p,e}$ in
order to provide an implicit representation of the domains' set
$(\mathcal{D}_{s})$, the players' set $N_{e}$, the functions
$(\mathcal{F}_{f,e})$ and the relations $(\mathcal{P}_{p,e})$,
respectively. Thus, we only evaluate them on demand.

The algorithm is similar to the CTL model checking algorithm
\cite{SMV} that operates by labelling each state $e\in\mathcal{S}E$
with the set of $labels(e)$ of sub-formulas of $\alpha$ which are
true in $e$. The algorithm starts with the set $labels(e)$ as the
empty set\footnote{The CTL model checking algorithm starts the set
of labels(e) as the set of propositions in $e$. In our algorithm we
just evaluate the predicates and functions on demand.} and then goes
by a series of steps (the number of operators in $\alpha$). At each
step $k$, sub-formulas with $k-1$ nested GAL operators are
processed. When a formula is processed, it is added to the labelling
of the state in which it is true. Thus,
$\mathcal{G},(\sigma_{s})\models_{e}\alpha\Longleftrightarrow\alpha\in
labels(e)$.

As GAL-formulas are represented in terms of $i$,
$p(t^{1}_{s_{1}},...,t_{s_{n}}^{n})$,
$(t^{1}_{s_{1}}$$\approx$$t_{s_{1}}^{\prime})$, $(\lnot\alpha)$,
$(\alpha\rightarrow\beta)$, $\exists x_{s_k}\alpha$, $[AX]\alpha$,
$E(\alpha\mathcal{U}\beta)$, $A(\alpha\mathcal{U}\beta)$, it is
sufficient to handle these cases. The cases $(\lnot\alpha)$,
$(\alpha\rightarrow\beta)$, $[AX]\alpha$,
$E(\alpha\mathcal{U}\beta)$ and $A(\alpha\mathcal{U}\beta)$ are
similar to the CTL model checking algorithm and we do not present
here (see \cite{Mc93Thesis} for more details). Below we present and
give the time complexity of the other procedures. In order to
guarantee termination of the algorithm, the functions
$(\mathcal{F}_{f,e})$ and the relations $(\mathcal{P}_{p,e})$ must
terminate since the model is finite this is accomplished. We use the
notation $\bar{\sigma}_{s_{1}}(e,t^{1}_{s_{1}})$ as the function
that interprets the term $t^{1}_{s_{1}}$ at the state~$e$. We take
its complexity as an upper bound on the implementation of
$\bar{\sigma}_{s_{1}}$ taking all states into account. We refer to
this upper bound as $|\bar{\sigma}_{s_{1}}(e,t^{1}_{s_{1}})|$.
\begin{itemize}
\item Case $i$:
The procedure \emph{verifyPlayer} (see Algorithm \ref{procPlayer})
labels all states $e\in\mathcal{S}E$ with the player $i$ if the
player $i$ belongs to the set of players in $e$. This procedure
requires time $O(|\mathcal{S}E|)$.
\item Case $p(t^{1}_{s_{1}},...,t^{n}_{s_{n}})$: The procedure
\emph{verifyPredicate} (see Algorithm \ref{procPred}) labels all
states $e\in\mathcal{S}E$ in which the interpretation of the
predicate $p$ with the interpretation of terms
$t^{1}_{s_{1}},...,t^{n}_{s_{n}}$ is true in $e$. This procedure
requires time
$O((|\bar{\sigma}_{s_{1}}(e,t^{1}_{s_{1}})|+...+|\bar{\sigma}_{s_{n}}
(e,t^{n}_{s_{n}})|)\times |\mathcal{S}E|)$. \footnote{Notice that
the evaluation of the terms and the predicate are done in all states
and the time complexity of them could not be polynomial.}
\item Case $t^{1}_{s_{1}}$$\approx$$t^{\prime}_{s_{1}}$: The
procedure \emph{verifyEquality} (see Algorithm \ref{procEqual})
labels all state $e\in\mathcal{S}E$ in which the interpretation of
the terms $t^{1}_{s_{1}}$ and $t^{\prime}_{s_{1}}$ are equal. The
time complexity is
$O((|\bar{\sigma}_{s_{1}}(e,t^{1}_{s_{1}})|+|\bar{\sigma}_{s_{n}}
(e,t^{\prime}_{s_{1}})|)\times |\mathcal{S}E|)$.
\item The
procedure \emph{verifyExists} (see Algorithm \ref{procExists})
labels all states $e\in\mathcal{S}E$ in which the formula $\alpha$
with all occurrences of the variable $x_{s_{k}}$ substituted by at
least one element of the domain is true. We use the notation
$\alpha[x_{s_k}\leftarrow d]$ as a function that substitutes all
occurrence of $x_{s_k}$ by $d$ in $\alpha$. This procedure requires
$O(|\mathcal{D}_{s_{k}}|\times |\mathcal{S}E|)$.
\end{itemize}

Thus, the complexity of the algorithm regards to: 1- The size of the
domains' set; 2- The size of the states' set; 3- The size of the
actions' set; 4- The complexity of both functions and predicates in
each state.
\begin{algorithm}
\caption{procedure verifyPlayer(i)} \label{procPlayer}
\begin{algorithmic}
\FORALL { $e\in\mathcal{S}E$}
    \IF{$i\in \mathcal{N}(e)$}
        \STATE $label(e):=label(e)\cup\{i\}$
    \ENDIF
\ENDFOR
\end{algorithmic}
\end{algorithm} \begin{algorithm}
\caption{procedure
verifyPredicate($p(t^{1}_{s_{1}},...,t^{n}_{s_{n}}))$}
\label{procPred}
\begin{algorithmic}
\FORALL { $e\in\mathcal{S}E$}
    \IF{$\langle\bar{\sigma}_{s_{1}}(e,t^{1}_{s_{1}}),...,\bar{\sigma}_{s_{n}}
(e,t^{n}_{s_{n}})\rangle \in \mathcal{P}(p,e)$}
        \STATE $label(e):=label(e)\cup\{p(t^{1}_{s_{1}},...,t^{n}_{s_{n}})\}$
    \ENDIF
\ENDFOR
\end{algorithmic}
\end{algorithm}
\begin{algorithm}
\caption{procedure verifyEquality$(t^{1}_{s_{1}}\approx
t^{\prime}_{s_{1}})$} \label{procEqual}
\begin{algorithmic}
\FORALL { $e\in\mathcal{S}E$}
    \IF{$\bar{\sigma}_{s_{1}}(e,t^{1}_{s_{1}})=\bar{\sigma}_{s_{1}}(e,t^{\prime}_{s_{1}})$}
        \STATE $label(e):=label(e)\cup\{t^{1}_{s_{1}}\approx t^{\prime}_{s_{1}}\}$
    \ENDIF
\ENDFOR
\end{algorithmic}
\end{algorithm}
\begin{algorithm}
\caption{procedure verifyExists$(\exists x_{s_{k}}\alpha)$}
\label{procExists}
\begin{algorithmic}
\FORALL { $d\in\mathcal{D}({s_{k}})$} \STATE $T := \{e \textrm{ } |
\textrm{ } \alpha[x_{s_{k}}\leftarrow d]\in label(e)\}$
 \FORALL { $e\in T$}
     \IF{$\exists x_{s_{k}}\alpha \not\in
label(e)$}
        \STATE $label(e):=label(e)\cup\{\exists
x_{s_{k}}\alpha\}$
    \ENDIF
\ENDFOR \ENDFOR
\end{algorithmic}
\end{algorithm}

Below we consider a simpler version of GAL, for the sake of a
simpler presentation of the algorithm's complexity. The non-logical
language has only one sort $D$ and one unary predicate $p:D$. Let
$\mathcal{G}_{S}$ be a GAL-structure, where: 1- The predicate $p$ is
interpreted as constant for all states and its time complexity is
represented by $O(p)$; 2- The size of sort D's domain is
$|\mathcal{D}|$; 3- The size of the states' set is $|\mathcal{S}E|$;
4- The size of the actions' set is $|\mathcal{CA}|$. Let $\alpha$ be
a GAL-formula for this language, where $\alpha_{M}$ and $\alpha_{D}$
are the number of modal connectives and the number of quantifier
connectives, respectively, in the formula $\alpha$. The time
complexity to verify $\alpha$ for $\mathcal{G}_{S}$ is
\[O(|\mathcal{D}|^{\alpha_{D}}\times |\alpha_{M}|\times((|\mathcal{S}E|\times
O(p))+|\mathcal{CA}|))\]

We have a prototype, namely Game Analysis Logic Verifier (GALV),
that was written as \emph{framework} in Java. GALV is available for
download at http://www.tecmf.inf.puc-rio.br/DaviRomero. All of the
examples that we will show in this article are implemented in our
prototype. The main advantages of this model checker are: 1- It
allows the use of abstract data types, for example, a list can be
used to represent the history of the game; 2- It might use a large
sort of libraries that are available in Java; 3- Functions and
predicates might be used to analyze games, such as the evaluation
functions that are used in the AI community to provide an estimate
of the expected utility of the game from a given position; 4- GALV
allows computational aspects to define the players' actions, for
example, a \emph{minimax} algorithm can be used to define the
actions of a certain player, while the other players might use
different algorithms. So, the time complexity to generate a game
might not be polynomial, i.e., it depends on the algorithms that
have been used to define the players' actions.

\section{Game Theory in Game Analysis Logic}\label{sectionGTinGAL}
We can model both the standard models and the standard solution
concepts of Game Theory using GAL. In this section we show that the
standard models are related to as GAL-structures and the standard
solution concepts are related to as GAL-formulas. Precisely, we
present the correspondence between the extensive games and the
GAL-structures as well as the solution concepts of Nash equilibrium
(NE) and subgame perfect equilibrium (SPE) and the formulas of GAL.
For more details about the rationale of the
definitions related to Game Theory see \cite{Games}. In the
sequel, we write down the definitions and theorems used in this
article.

\begin{figure}[h]
\setlength{\unitlength}{1 cm}
\begin{picture}(4,5.0)

 \put(3.3,0.8){$0$,$0$}
 \put(4.9,0.8){$2$,$1$}
 \put(6.5,2.2){$1$,$2$}

 \put(4.3,3.2){A}
 \put(6.3,3.2){B}
 \put(3.5,1.7){L}
 \put(4.7,1.7){R}

 \put(5.5,4){\circle{0.5}}
 \put(5.4,3.9){1}
\put(5.5,3.75){\vector(1,-1){1.2}}
\put(5.5,3.75){\vector(-1,-1){1.2}}

 \put(4.3,2.3){\circle{0.5}}
 \put(4.2,2.2){2}
 \put(4.3,2.05){\vector(1,-1){0.8}}
 \put(4.3,2.05){\vector(-1,-1){0.8}}

 \put(2,0){(a) - Extensive form representation}
 \put(8.3,0){(b) - A GAL representation}

 \put(11.5,4.5){\oval(1.7,1)}
 \put(11.1,4.6){$h=\emptyset$}
 \put(11.3,4.2){$\{1\}$}
 \put(11.5,4.0){\vector(-2,-1){1}}
 \put(11.5,4.0){\vector(2,-1){1}}

 \put(10.5,3){\oval(2.0,1)}
 \put(9.8,3.1){$h=(A)$}
 \put(10.3,2.7){$\{2\}$}
 \put(10.5,2.5){\vector(-3,-1){1.2}}
 \put(10.5,2.5){\vector(3,-1){1.2}}

 \put(9.2,1.5){\oval(2.2,1.1)}
 \put(8.3,1.6){$h=(A,L)$}
 \put(9.1,1.2){$\{\}$}

 \put(11.5,1.5){\oval(2.2,1.1)}
 \put(10.6,1.6){$h=(A,R)$}
 \put(11.4,1.2){$\{\}$}

 \put(12.6,3){\oval(2.0,1)}
 \put(11.9,3.1){$h=(B)$}
 \put(12.4,2.7){$\{\}$}
\end{picture}
\caption{\emph{Mapping an extensive game into a GAL model.
}}\label{ExtensiveGameFigure}
\end{figure}
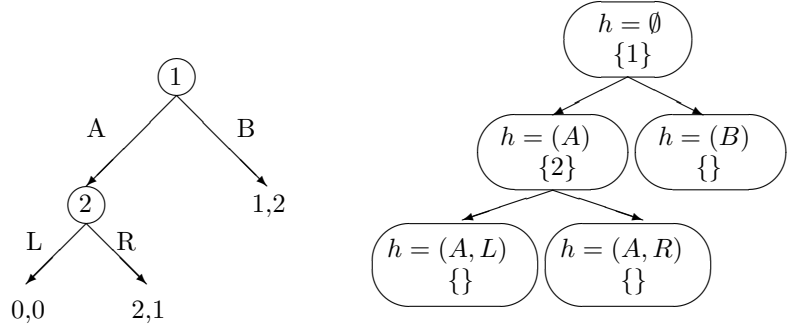

An extensive game is a model in which each player can consider his
or her plan of action at every time of the game at which he or she
has to make a choice. There are two kinds of models: game with
perfect information; and games with imperfect information. For the
sake of simplicity we restrict the games to models of perfect
information. A general model that allows imperfect information is
straightforward. Below we present the formal definition and the
example depicted in Figure \ref{ExtensiveGameFigure}.a.

\begin{definition}\label{extensiveDefinition}
An \textbf{extensive game with perfect information} is a tuple
$\langle
\textbf{N},\textbf{H},\textbf{P},(\textbf{u$_{\textbf{i}}$})
\rangle$, where
\begin{itemize}
\item \textbf{N} is a set, called the set of players.
\item \textbf{H} is a set of sequences of actions (finite or infinite),
called the set of histories, that satisfies the following properties
\begin{itemize}
\item the empty sequence is a history, i.e. $\emptyset\in H$.
\item if $(a_{k})_{k\in K}\in H$ where $K\subseteq\mathbb{N}$ and for all $l\leq |K|$, then $(a_{k})_{k=0,\ldots,l}\in
H$.
\item if $(a_{0}\ldots a_{k})\in H$ for all $k\in\mathbb{N}$, then
the infinite sequence $(a_{0}a_{1}\ldots)\in H$.
\end{itemize}
A history $h$ is
\textbf{terminal} if it is infinite or it has no action $a$ such
that $(h,a)\in H$. We refer to $\textbf{T}$ as the set of terminals.
\item $\textbf{P}$ is a function that assigns to each non-terminal history a
player.
\item For each player $i\in N$, a utility function $\textbf{u$_{\textbf{i}}$}$
on $T$.
\end{itemize}
\end{definition}

\begin{example}\label{extensiveGameExample}
An example of a two-player extensive game $\langle
\textbf{N},\textbf{H},\textbf{P},(\textbf{u$_{\textbf{i}}$})
\rangle$, where:
\begin{itemize}
\item $\textbf{N}=\{1,2\}$;
\item $\textbf{H}=\{\emptyset,(A),(B),(A,L),(A,R)\}$;
\item $\textbf{P}(\emptyset)=1$ and $\textbf{P}((A))=2$;
\item $\textbf{u$_{\textbf{1}}$}((B))=1,$ $\textbf{u$_{\textbf{1}}$}((A,L))=0,$ $\textbf{u$_{\textbf{1}}$}((A,R))=2$;
\item $\textbf{u$_{\textbf{2}}$}((B))=2,\textbf{u$_{\textbf{2}}$}((A,L))=0,\textbf{u$_{\textbf{2}}$}((A,R))=1$.
\end{itemize}
\end{example}

A \emph{strategy of player i} is a function that assigns an action
for each non-terminal history for each $P(h)=i$. For the purpose of
this article, we represent a strategy as a tuple. In order to avoid
confusing when we refer to the strategies or the histories, we use
`$\langle$' and `$\rangle$' to the strategies and `$($' and `$)$' to
the histories. In Example \ref{extensiveGameExample}, player $1$ has
to make a decision only after the initial state and he or she has
two strategies $\langle A\rangle$ and $\langle B\rangle$. Player $2$
has to make a decision after the history $(A)$ and he or she has two
strategies $\langle L\rangle$ and $\langle R\rangle$. We denote
$\textbf{S}_{\textbf{i}}$ as the set of player i's strategies. We
denote $s=(s_{i})$ as a \textbf{strategy profile}. We refer to
$\textbf{O(s}_{\textbf{1}},\ldots,\textbf{s}_{\textbf{n}}\textbf{)}$
as an outcome that is the terminal history when each player follows
his or her strategy $s_{i}$. In Example \ref{extensiveGameExample},
$\langle \langle B\rangle,\langle L\rangle\rangle$ is a strategy
profile in which the player $1$ chooses $B$ after the initial state
and the player 2 chooses $L$ after the history $(A)$, and $O(\langle
B\rangle,\langle L\rangle)$ is the outcome $(B)$. In a similar way,
we refer to
$\textbf{O}_{\textbf{h}}\textbf{(h,}\textbf{s}_{\textbf{1}},\ldots,\textbf{s}_{\textbf{n}}\textbf{)}$
as the outcome when each player follows his or her strategy $s_{i}$
from history $\textbf{h}$. In Example \ref{extensiveGameExample},
$O_h((A),\langle B\rangle,\langle L\rangle)$ is the outcome $(A,L)$
and $u_1((A),\langle B\rangle,\langle L\rangle)=u_1((A,L))=0$.

We can model an extensive game $\Gamma=\langle N,H,P,(u_{i})\rangle$
as a GAL-structure in the following way. Each history $h\in H$ (from
the extensive game) is represented by a state, in which a 0-ary
symbol $h$ designates a history of $\Gamma$ (the one that the state
is coming from), so $h$ is a non-rigid designator. The set of the
actions of the GAL-structure is determined by the set of actions of
each history, i.e., given a history $h\in H$ and an action $a$ such
that $(h,a)\in H$, then the states namely $h$ and $(h,a)$ are in the
set of actions of the GAL-structure, i.e. $\langle
h,(h,a)\rangle\in\mathcal{CA}$. Function $P$ determines the player
that has to make a choice at every state, i.e. $N_{h}=\{P(h)\}$. The
utilities functions are rigidly defined as in the extensive game.
The initial state is the state represented by the initial history of
the extensive game, i.e. $H_o=\{\emptyset\}$. Sorts $H$ and $T$ are
interpreted as the histories and terminal histories of the extensive
game, respectively, i.e., $\mathcal{D}_{H}=H$ and
$\mathcal{D}_{T}=T$. Sort $U$ represents the utility values and is
interpreted as the set of all possible utility values of the
extensive game\footnote{Note that this set is finite if the game is
finite.}. In order to define the solution concept of the subgame
perfect equilibrium and the Nash equilibrium, we add to this
structure the sets of players' strategies $(\mathcal{D}_{S_{i}})$
and functions $O$ and $O_h$. To summarize, a \textbf{GAL-structure
for an extensive game with perfect information} $\Gamma =\langle
N,P,H,(u_{i})\rangle$ is the tuple $\langle
H,H_{o},\mathcal{CA},\textbf{(}\mathcal{D}_{H},\mathcal{D}_{T},\mathcal{D}_{S_{i}},\mathcal{D}_{U}\textbf{)},
\textbf{(}u_{i},h_{h},O,O_h\textbf{)},~(\geq)~,(N_{h})\rangle$ with
non-logic language $\langle (H,T,S_i,U)$ $,(h:\rightarrow
H,u_i:T\rightarrow U,O:S\rightarrow T,O_h:H\times S\rightarrow T)$
$,(\geq:U\times U),N\rangle$. The example below is the GAL-structure
(see Figure \ref{ExtensiveGameFigure}.b) of Example
\ref{extensiveGameExample} (see Figure \ref{ExtensiveGameFigure}.a).

\begin{example}\label{exampleExtenGameGal2}
The GAL-structure of Example \ref{extensiveGameExample} is $\langle
H,H_{o},\mathcal{CA},$ $\textbf{(}\mathcal{D}_{H},\mathcal{D}_{T},\mathcal{D}_{S_{1}},\mathcal{D}_{S_{2}},\mathcal{D}_{U}\textbf{)},\textbf{(}h_{h},u_{1},u_{2},$
$O,O_h\textbf{)},(\geq),\textbf{(}N_{h}\textbf{)}\rangle$
   with non-logic language $\langle(H,T,S_1,S_2,U),(h:\rightarrow H,u_{1}:T\rightarrow U,
   u_{2}:T\rightarrow U,O:S_1\times S_2\rightarrow T,$ $O_h:H\times S_1\times S_2\rightarrow T),(\geq:U \times U),\{1,2\}\rangle$
where
\begin{itemize}
\item
$H=\{\emptyset,~(A),~(B),~(A,L),~(A,R)\}$ and $H_{o}=\{\emptyset\}$.
\item $\mathcal{CA}=\{\langle\emptyset,~(A)\rangle,~\langle\emptyset,(B)\rangle,~\langle(A),(A,L)\rangle,~\langle(A),(A,R)\rangle\}$.
\item $\mathcal{D}_{S_{1}}=\{\langle A\rangle,\langle B\rangle\}$, $\mathcal{D}_{S_{2}}=\{\langle L\rangle,\langle R\rangle\}$ and $\mathcal{D}_{U}=\{0,1,2\}$.
\item $\mathcal{D}_{H}=\{\emptyset,~(A),~(B),~(A,L),~(A,R)\}$ and $\mathcal{D}_T=\{(B),~(A,L),~(A,R)\}$.
\item $h_{\emptyset}=\emptyset$, $h_{(A)}=(A)$, $h_{(B)}=(B)$, $h_{(A,L)}=(A,L)$,
$h_{(A,R)}=(A,R)$.
\item $N_{\emptyset}=\{1\}$, $N_{(A)}=\{2\}$,
$N_{(B)}=N_{(A,L)}=N_{(A,R)}=\{\}$.
\item Functions $O$, $O_h$, $u_1$ and $u_2$ are rigidly defined as in the
extensive game.
\end{itemize}
\end{example}

The most used solution concepts for extensive games are Nash
equilibrium (NE) and subgame perfect equilibrium (SPE). The solution
concept of NE requires that each player's strategy be optimal, given
the other players' strategies. And, the solution concept of SPE
requires that the action prescribed by each player's strategy be
optimal, given the other players' strategies, after every history.
In SPE concept, the structure of the extensive game is taken into
account explicitly, while, in the solution concept of NE, the
structure is taken into account only implicity in the definition of
the strategies. Below we present the SPE definition in a standard
way. The NE definition below regards to the structure of an
extensive game, yet is an equivalent one to the standard.

\begin{definition}\label{defSubgame1}
A \textbf{subgame perfect equilibrium (SPE)} of an extensive game
$\Gamma=\langle N,H,P,(u_i)\rangle$ is a strategy profile
$s^{*}=\langle s_1^{*},\ldots,s_n^{*}\rangle$ such that for every
player $i\in N$ and every history $h\in H$ for which $P(h)=i$ we
have
\[u_i(O_h(h,s^*_{1},\ldots,s^*_n))\geq
u_i(O_h(h,s^*_{1},\ldots,s_i,\ldots,s^*_n)),\] for every strategy
$s_{i}\in S_{i}$.
\end{definition}

\begin{definition}\label{defNash1}
A \textbf{Nash equilibrium (NE)} of an extensive game
$\Gamma=\langle N,H,P,(u_i)\rangle$ is a strategy profile
$s^*=\langle s_1^{*},\ldots,s_n^{*}\rangle$ such that for every
player $i\in N$ and every history on the path of the strategy
profile $s^*$ (i.e. $h\in O(s^*)$) for which $P(h)=i$ we have
\[u_i(O_h(h,s^*_{1},\ldots,s^*_n))\geq
u_i(O_h(h,s^*_{1},\ldots,s_i,\ldots,s^*_n)),\] for every strategy
$s_i\in S_{i}$.
\end{definition}

We invite the reader to verify that the strategy profiles $\langle
\langle A\rangle ,\langle R\rangle\rangle$ and $\langle \langle
B\rangle ,\langle L\rangle\rangle$ are the Nash equilibria in
Example \ref{extensiveGameExample}. Game theorists can argue that
the solution $\langle \langle B\rangle ,\langle L\rangle\rangle$ is
not reasonable when the players regard to the sequence of the
actions. To see that the reader must observe that after the history
$(A)$ there is no way for player 2 commit himself or herself to
choose $L$ instead of $R$ since he or she will be better off
choosing $R$ (his or her utility is 1 instead of 0). Thus, player 2
has an incentive to deviate from the equilibrium, so this solution
is not a subgame perfect equilibrium. On the other hand, we invite
the reader to verify that the solution $\langle \langle A\rangle
,\langle R\rangle\rangle$ is the only subgame perfect equilibrium.

Consider formulas \ref{formSPE1} and \ref{formNash1} as expressing
subgame perfect equilibrium definition \ref{defSubgame1} and Nash
equilibrium definition \ref{defNash1}, respectively. A strategy
profile $s^{*}=\langle s_1^{*},\ldots,s_n^{*}\rangle$ is a SPE (or
NE) if and only if formula \ref{formSPE1} (or formula
\ref{formNash1}) holds at the initial state $\emptyset$, where each
$\sigma_{S_i}(v_{s_i}^{*})=s_{i}^{*}$.
\begin{small}
\begin{equation}[AG]\left({\textstyle\bigwedge\limits_{i\in N}}
i\rightarrow\forall v_{s_{i}}
\left(u_{i}(O_h(h,v_{s_{1}}^{*},\ldots,v_{s_{n}}^{*}))\geq
u_{i}(O_h(h,v_{s_{1}}^{*},\ldots,v_{s_{i}},\ldots,v_{s_{n}}^{*}))\right)\right)\label{formSPE1}
\end{equation}

\begin{equation}[EG]\left(
    \begin{array}{c}
            h\in O(v_{s_{1}}^{*},\ldots,v_{s_{n}}^{*})~~~~\wedge
            \\ \left({\textstyle\bigwedge\limits_{i\in N}} i\rightarrow\forall
v_{s_{i}} \left(u_{i}(O_h(h,v_{s_{1}}^{*},\ldots,v_{s_{n}}^{*}))\geq
u_{i}(O_h(h,(v_{s_{1}}^{*},\ldots,v_{s_{i}},\ldots,v_{s_{n}}^{*})))\right)\right)\end{array}\right)\label{formNash1}
\end{equation}
\end{small}

In order to guarantee the correctness of the representation of both
subgame perfect equilibrium and Nash equilibrium, we state the
theorem below. The proof is provided in Appendix \ref{appendix}.
\begin{theorem}\label{teorema}
Let $\Gamma$ be an extensive game, and $\mathcal{G}_{\Gamma}$ be a
GAL-structure for $\Gamma$, and $\alpha$ be a subgame perfect
equilibrium formula for $\mathcal{G}$ as defined in Equation
\ref{formSPE1}, and $\beta$ be a Nash equilibrium formula as defined
in Equation \ref{formNash1}, and $(s_{i}^{*})$ be a strategy
profile, and $(\sigma_{S_{i}})$ be valuations functions for sorts
$(S_{i})$.
\begin{itemize}
\item $\textrm{A strategy profile } (s_{i}^{*}) \textrm{ is a SPE of }\Gamma\Longleftrightarrow
\mathcal{G}_{\Gamma},\!(\sigma_{S_i}\!)\!\!\models_{\emptyset}\alpha$, where
each $\sigma_{S_i}(v_{s_{i}}^{*})=s_{i}^{*}$
\item $\textrm{A strategy profile } (s_{i}^{*}) \textrm{ is a NE of }\Gamma
\Longleftrightarrow\mathcal{G}_{\Gamma},\!(\sigma_{S_i}\!)\!\!\models_{\emptyset}\beta$,
where each $\sigma_{S_i}(v_{s_{i}}^{*})=s_{i}^{*}$
\end{itemize}
\end{theorem}

\section{Experimental Results}\label{sectionExper}
In this section we show the performance of the GAL model checking
algorithm against other algorithms. The algorithm was written in
Java and the experiments were executed on a 2.4GHz Celeron with 512
MBytes of RAM, running Windows XP Home Edition.

\begin{figure}[h]
 \begin{tabular}[l]{cc}
  \raisebox{-0pt}{
      \includegraphics[width=.45\textwidth]{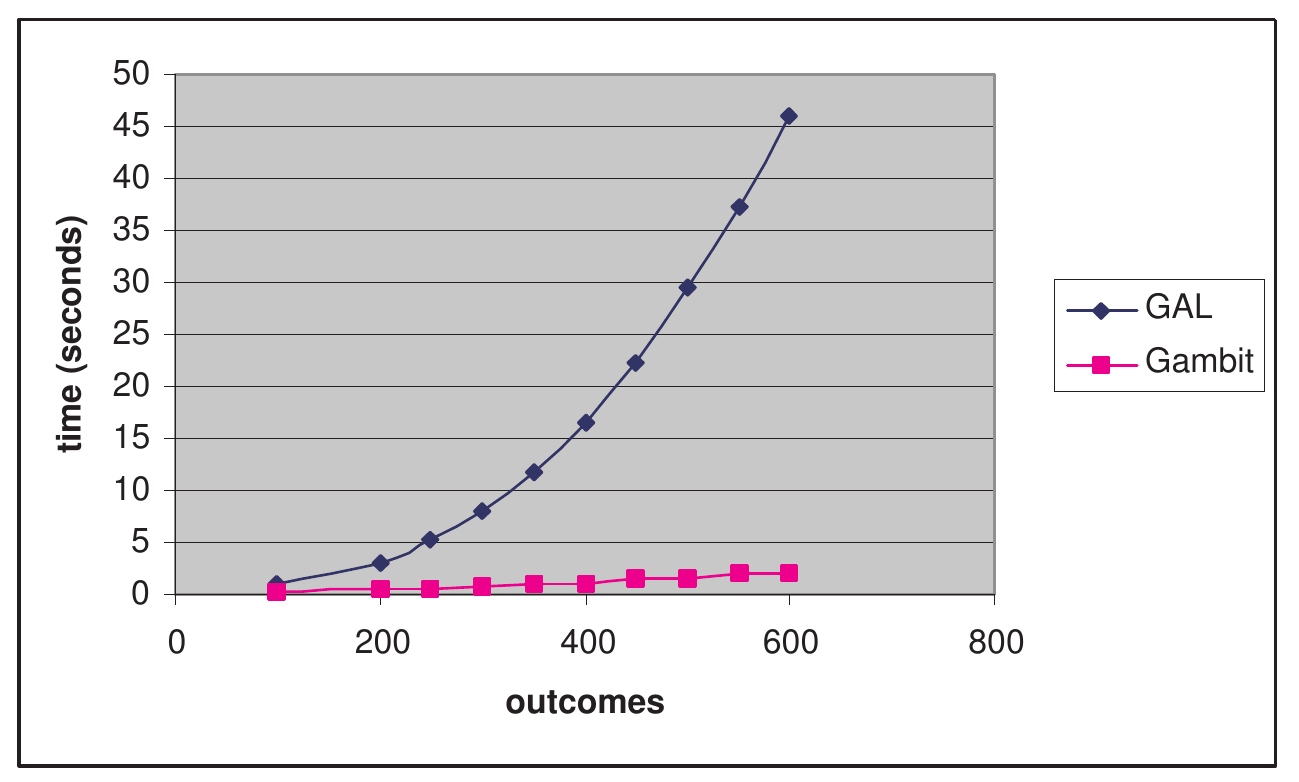}
      }
  &
  \raisebox{-0pt}{
      \includegraphics[width=.45\textwidth]{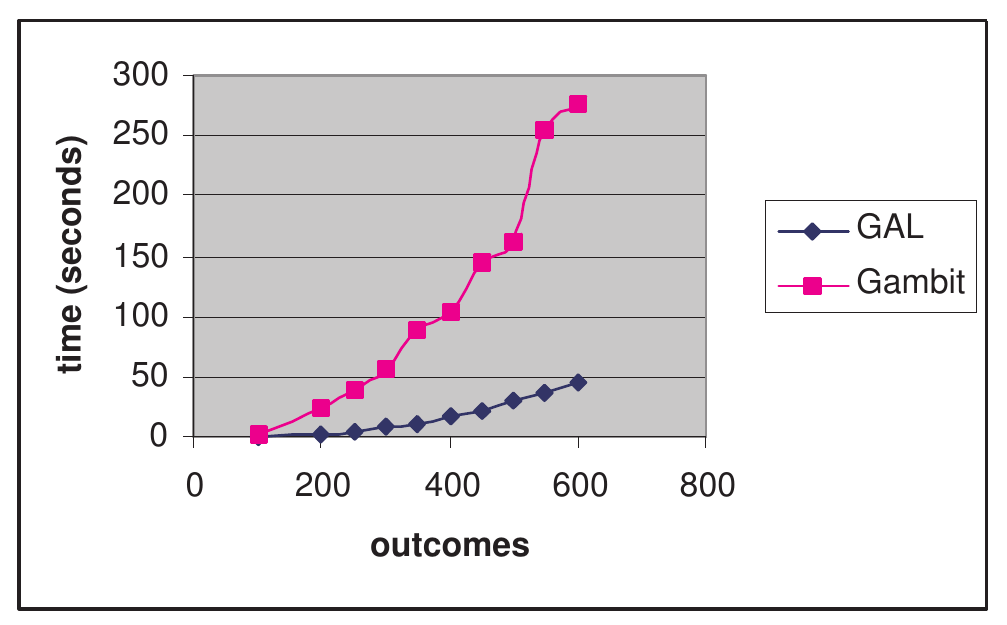}
    }
      \\ (a) - Randomized payoffs & (b) - Constant payoffs
\end{tabular}
\caption{Two-player games.}\label{figStratGame}
\end{figure}

Several algorithms for the problem of finding a Nash equilibrium are
proposed in the literature (see \cite{mckelvey96computation} for a
survey). Most of them compute a mixed Nash equilibrium. Gambit
\cite{GambitManual} is the best-known Game Theory software that
implements most of all algorithms. We use both Gambit (with its
\emph{EnumPureSolve} method) and our algorithm in order to compute
the pure Nash Equilibria. Figure \ref{figStratGame} shows the
running times (in seconds) of several two-player games in which the
payoffs of the games were randomly generated (Figure
\ref{figStratGame}.a) or were taken as the constant value 0 (Figure
\ref{figStratGame}.b). The difference between the games in Figure
\ref{figStratGame}.a and Figure \ref{figStratGame}.b relies on the
size of the set of equilibria. Our algorithm took almost the same
time to find the solution concept regardless of the size of
equilibria. On the other hand, Gambit's performance was much more
dependent on the size of equilibria as shown in Figure
\ref{figStratGame}.

\noindent

In \cite{Vasconcelos03,Vasconcelos03Laptec} is proposed a
metalanguage to describe games, namely \emph{RollGame}, as well as a
translation into the input language of the well-known SMV model
checker \cite{Mc93Thesis}, in order to reason about games. In this
section, we take Tic-Tac-Toe game in order to provide an example
that an explicit representation of such a game can be more efficient
than using an OBDD approach as in SMV. It is worth mentioning that
SMV uses a propositional logic (CTL), so it cannot express many
solution concepts as defined in Section \ref{sectionGTinGAL}.
Moreover, it does not allow the use of abstract data types, yet the
usage of integer is prohibited in many situations, such as when one
wants to use utilities values.

In \cite{Vasconcelos03,Vasconcelos03Laptec}, a version of a
Tic-Tac-Toe game is modeled and analyzed. In this version, one of
the players (PlayerX) uses a certain strategy, while the other
player (PlayerO) spreads all possible actions. It is also shown that
the strategy of PlayerX never reaches a losing position in the game.
This property is expressed by the CTL formula defined in Equation
\ref{formAFWinX} below, which states that PlayerX will always win or
draw. We also model this game with the same strategy using our
algorithm, and the performance of verifying this formula is much
better in our algorithm (0.001 seconds) than using the SMV model
checker (45.211 seconds). However, we should also take into account
the time to generate this game in order to compare our algorithm
with SMV. The required time was 0.289 seconds; so our algorithm took
0.290 seconds to generate and analyze this version of Tic-Tac-Toe
game\footnote{Here, we refer to the average (arithmetic mean) time
of 10 runs of each approach. The standard deviation with SMV and our
algorithm were 1.333 and 0.009, respectively.}.
\begin{equation}
[AF](winX \vee Draw)\label{formAFWinX}
\end{equation}
As we have claimed at the end of Section \ref{sectionGALV}, one of
the main advantages of the GALV model checker is that it allows
computational aspects in the modeling language. Thus, we are able to
use standard algorithms of the AI community to model and analyze a
game. We take Tic-Tac-Toe as an example again, and we define one of
the players (PlayerX), using a \emph{minimax} algorithm with maximal
depth (9), while the other player (PlayerO) spreads all the possible
actions. The required time for generate the game was 14.718 seconds
and to analyze the GAL formula defined in Equation \ref{formAFWinX}
was 0.001 seconds. Note that this approach is not possible using a
standard model checker, such as SMV or SPIN.


\section{Conclusion and Future Works}

In this work, we have presented a first-order modal logic (GAL) to
model and analyze games. We have also provided a model checking
algorithm for GAL to achieve automatic verification for finite
games. We have illustrated in Section \ref{sectionGTinGAL} that
standard concepts of Game Theory can be modeled in GAL. Using our
prototype of a GAL model checker, we have performed case studies in
at least two directions: as a tool to find solution concepts of Game
Theory; and as a tool to analyze games that are based on standard
algorithms of the AI community, such as \emph{minimax} algorithm.
Despite the fact that our algorithm uses an explicit representation,
it outperforms the SMV model-checker as shown in Section
\ref{sectionExper}. This might suggest that an explicit
representation is better for games than using a symbolic
representation as OBDD. However, a general conclusion cannot be
drawn. Some future works are still needed and are listed below
\begin{itemize}
\item Define an adequate and sound system of GAL that is able to prove
formal theorems of Game Theory, such as the existence of mixed Nash
equilibrium in strategic games.
\item Implement a player of a game using formulas of GAL, such as the subgame perfect
equilibrium formula as shown in Section \ref{sectionGTinGAL}. This
approach might use evaluation functions and be limited to a certain
depth as in a \emph{minimax} procedure. As this is an heuristic
approach, we argue that define other solution concepts in a logic
framework is easier than to implement new algorithms. For instance,
we can define the strategy of a player according to a conjunction of
the subgame perfect formula and a Pareto Optimal formula.
\item Improve
the performance of the GAL model checker, since it uses an explicit
representation. We cannot use an OBDD-approach since in GAL we are
dealing with a first-order interpretation that may vary over the
states of the game.
\end{itemize}

\newif\ifabfull\abfulltrue

\appendix
\section{Proof of Theorem \ref{teorema}}\label{appendix}
\begin{theorem}
Let $\Gamma$ be an extensive game, and $\mathcal{G}_{\Gamma}$ be a
GAL-structure for $\Gamma$, and $\alpha$ be a subgame perfect
equilibrium formula for $\mathcal{G}$ as defined in Equation
\ref{formSPE1}, and $\beta$ be a Nash equilibrium formula as defined
in Equation \ref{formNash1}, and $(s_{i}^{*})$ be a strategy
profile, and $(\sigma_{S_{i}})$ be valuations functions for sorts
$(S_{i})$.
\begin{itemize}
\item $\textrm{A strategy profile } (s_{i}^{*}) \textrm{ is a
SPE of }\Gamma\Longleftrightarrow
\mathcal{G}_{\Gamma},\!(\sigma_{S_i}\!)\!\!\!\models_{\emptyset}\alpha$, where
each $\sigma_{S_i}(v_{s_{i}}^{*})=s_{i}^{*}$
\item $\textrm{A
strategy profile } (s_{i}^{*}) \textrm{ is a NE of }\Gamma
\Longleftrightarrow\mathcal{G}_{\Gamma},\!(\sigma_{S_i}\!)\!\!\!\models_{\emptyset}\beta$,
where each $\sigma_{S_i}(v_{s_{i}}^{*})=s_{i}^{*}$
\end{itemize}
\begin{proof}
\begin{itemize}
\item
A strategy profile $(s_{i}^{*})$ is a SPE of $\Gamma$
$\Longleftrightarrow\mathcal{G}_{\Gamma},(\sigma_{S_i})\models_{\emptyset}\alpha$,
where each $\sigma_{S_i}(v_{s_{i}}^{*})=s_{i}^{*}$.
\\A strategy profile $(s_{i}^{*})$ is a SPE of $\Gamma$.
\\ $\Longleftrightarrow$ for every player $i$ and every
history $h\in H$ for which $P(h)=i$ we have\\
$u_i(O_h(h,s^*_{1},\ldots,s^*_n))\geq
u_i(O_h(h,s^*_{1},\ldots,s_i,\ldots,s^*_n))$, for every strategy
$s_{i}\in S_{i}$.
\\ By the definition of
$\mathcal{G}_\Gamma$ from $\Gamma$, we have that every state of
$\mathcal{G}_\Gamma$, which represents a history of $\Gamma$, is
reached by a path from the initial state $\emptyset$; moreover, we
have that each domain of player i's strategy $\mathcal{D}_{S_i}$ is
interpreted by the set of strategies $S_i$ (i.e.
$\mathcal{D}_{S_i}=S_i$), and the player that has to take a move in
a state $e_k$, which represents the history $h_k$, is defined by the
function $P$ (i.e. $N_{e_k}=\{P(h_k)\}$), and, finally, the symbol
$h$ is interpreted in $e_k$ by the history $h_k$ (i.e.
$\bar{\sigma}_H(e_k,h)=h_k$). As a consequence of this definition,
we have
\\$\Longleftrightarrow$for all paths
$\pi(\emptyset)=e_0,e_1,\ldots$ and for all $k\geq 0$, for every
player $i\in N$ such that IF $i\in N_{e_k}$ THEN we have for all
$d_i\in\mathcal{D}_{S_i}$\\$\left(u_{i}(O_h(\bar{\sigma}_H(e_k,h),s^*_1,\ldots,s^*_n)\geq
u_{i}(O_h(\bar{\sigma}_H(e_k,h),s_{1}^{*},\ldots,d_{i},\ldots,s_{n}^{*}))\right)$.
\\ As function $O_h$ and utility functions $(u_i)$ are rigidly
interpreted as in the extensive game $\Gamma$, we have
\\$\Longleftrightarrow$ for all paths $\pi(\emptyset)=e_0,e_1,\ldots$ and for all $k\geq 0$, for
every player $i\in N$ such that IF
$\mathcal{G}_{\Gamma},(\sigma_{S_i})\models_{e_k}i$ THEN for all
$d_i\in\mathcal{D}_{S_i}$ we have
\\$\mathcal{G}_{\Gamma},(\sigma_{S_i}(v_{S_{i}}|d_i))\models_{e_k}\left(u_{i}(O_h(h,v_{S_{1}}^{*},\ldots,v_{S_{n}}^{*}))\geq
u_{i}(O_h(h,v_{S_{1}}^{*},\ldots,v_{S_{i}},\ldots,v_{S_{n}}^{*}))\right)$,\\
where $\sigma_{S_i}(v_{S_{i}}^{*})=s_{i}^{*}$.\\
$\Longleftrightarrow$ for all paths $\pi(\emptyset)=e_0,e_1,\ldots$
and for all $k\geq 0$ we have
\\$\mathcal{G}_{\Gamma},(\sigma_{S_i})\models_{e_k}\left({\textstyle\bigwedge\limits_{i\in
N}} i\rightarrow\forall v_{S_{i}}
\left(u_{i}(O_h(h,v_{S_{1}}^{*},\ldots,v_{S_{n}}^{*}))\geq
u_{i}(O_h(h,v_{S_{1}}^{*},\ldots,v_{S_{i}},\ldots,v_{S_{n}}^{*}))\right)\right)$,
\\ where each $\sigma_{S_i}(v_{S_{i}}^{*})=s_{i}^{*}$.
\\ $\Longleftrightarrow
\mathcal{G}_{\Gamma},\!(\sigma_{S_i})\!\!\models_{\emptyset}\![AG]\!\!\left({\textstyle\bigwedge\limits_{i\in
N}} i\rightarrow\forall v_{S_{i}}
\!\!\left(u_{i}(O_h(h,v_{S_{1}}^{*},\ldots,v_{S_{n}}^{*}))\geq
u_{i}(O_h(h,v_{S_{1}}^{*},\ldots,v_{S_{i}},\ldots,v_{S_{n}}^{*}))\!\right)\!\!\right)\!$,\\
where each $\sigma_{S_i}(v_{S_{i}}^{*})=s_{i}^{*}$.

\item
A strategy profile $(s_{i}^{*})$ is a NE of $\Gamma$
$\Longleftrightarrow\mathcal{G}_{\Gamma},(\sigma_{S_i})\models_{\emptyset}\beta$,
where each $\sigma_{S_i}(v_{s_{i}}^{*})=s_{i}^{*}$.
\\ A strategy profile $(s_{i}^{*})$ is a NE of $\Gamma$.
\\$\Longleftrightarrow$ for every player $i$ and every history
$h\in O(s^*)$ in which $P(h)=i$ we have
\\$u_i(O(s^*_{1},\ldots,s^*_n))\geq
u_i(O(s^*_{1},\ldots,s_i,\ldots,s^*_n))$, for every strategy
$s_{i}\in S_{i}$.
\\ We take the path
$\pi(\emptyset)=e_0,e_1,\ldots$ in $\mathcal{G}_{\Gamma}$ that is
defined by histories $h_0,h_1,\ldots$ on the equilibrium's path
$O(s^*_{1},\ldots,s^*_n)$ according to definition of
$\mathcal{G}_{\Gamma}$ from $\Gamma$. We have
\\ $\Longleftrightarrow$ there is a path
$\pi(\emptyset)=e_0,e_1,\ldots$ such that for all $k\geq 0$ we have
$\bar{\sigma}_H(e_k,h)\in
O(s_{1}^{*},\ldots,s_{n}^{*})$
\\ AND for every player $i\in N$ IF $i\in N_{e_k}$ THEN for all $s_i\in S_{i}$ we have
\\$\left(u_{i}(O_h(\bar{\sigma}_H(e_k,h),s_{1}^{*},\ldots,s_{n}^{*}))\geq
u_{i}(O_h(\bar{\sigma}_H(e_k,h),(s_{1}^{*},\ldots,s_{i},\ldots,s_{n}^{*})))\right)$,
\\ where each $\sigma_{S_{i}}(v_{S_{i}}^{*})=s_{i}^{*}$.
\\ As function $O,~O_h$ and utility functions $(u_i)$ are rigidly
interpreted as in the extensive game $\Gamma$, we have
\\$\Longleftrightarrow$ there is a path
$\pi(\emptyset)=e_0,e_1,\ldots$ such that for all $k\geq 0$ we have
\\$\mathcal{G}_{\Gamma},(\sigma_{S_{i}})\models_{e_k}h\in O(v_{S_{1}}^{*},\ldots,v_{S_{n}}^{*})$ AND
\\$\mathcal{G}_{\Gamma},(\sigma_{S_{i}})\models_{e_k}{\textstyle\bigwedge\limits_{i\in
N}} i\rightarrow\forall v_{S_{i}}
\left(u_{i}(O_h(h,v_{S_{1}}^{*},\ldots,v_{S_{n}}^{*}))\geq
u_{i}(O_h(h,(v_{S_{1}}^{*},\ldots,v_{S_{i}},\ldots,v_{S_{n}}^{*})))\right)$,
\\ where each $\sigma_{S_{i}}(v_{S_{i}}^{*})=s_{i}^{*}$.
\\ $\Longleftrightarrow$
\\$\!\!\!\!\!\!\mathcal{G}_{\Gamma},(\sigma_{S_{i}})\!\!\models_{\emptyset}\![EG]\!\!\left(
    \begin{array}{c}
            h\in O(v_{S_{1}}^{*},\ldots,v_{S_{n}}^{*})~~~~\wedge
            \\ \!\!\!\left({\textstyle\bigwedge\limits_{i\in N}} i\rightarrow\forall
v_{S_{i}} \left(u_{i}(O_h(h,v_{S_{1}}^{*},\ldots,v_{S_{n}}^{*}))\geq
u_{i}(O_h(h,(v_{S_{1}}^{*},\ldots,v_{S_{i}},\ldots,v_{S_{n}}^{*})))\right)\!\!\right)\end{array}\!\!\!\!\right)$,
\\ where each $\sigma_{S_{i}}(v_{S_{i}}^{*})=s_{i}^{*}$.
\end{itemize}
\end{proof}
\end{theorem}

\end{document}